\documentclass[useAMS,usenatbib]{mn2e}
\usepackage{graphics,graphicx}
\usepackage{color}
\usepackage{epsf}
\usepackage{epsfig}
\usepackage{dcolumn}

\newcommand{\lsim}{\raisebox{-.4ex}{$\stackrel{<}{\scriptstyle \sim}$}}
\newcommand{\msim}{\raisebox{-.4ex}{$\stackrel{>}{\scriptstyle \sim}$}}

\def \etal  {\hbox{et~al.\/}}
\def \ch    {\hbox{\em Chandra \/}}
 
\def \wr    {\hbox{WR\,65}}  
\def\changed{}

\usepackage{txfonts}
\title[New colliding wind binary WR\,65]{Puzzling X-rays from the 
new colliding wind binary \wr\ \changed{(WC9d)}}

\author[L. M. Oskinova \& W.-R. Hamann]
{L. M. Oskinova\thanks{E-mail: 
lida@astro.physik.uni-potsdam.de} \& W.-R. Hamann\\
Potsdam University,  Universit{\"a}tsstandort Golm,  Haus 28, Potsdam, 14476 
Germany}

\begin{document}

\date{Accepted 2008 July 29. Received 2008 July 25; in original 
form 2008 July 7}

\pagerange{\pageref{firstpage}--\pageref{lastpage}} \pubyear{2008}
\maketitle
\label{firstpage}

\begin{abstract} We report the discovery of variability in the X-ray 
emission from the Wolf-Rayet type star WR\,65. Using archival \ch\
data spanning over 5\,yr we detect changes of the X-ray flux by a
factor of 3 accompanied by changes in the X-ray spectra. We believe
that this X-ray emission originates from wind-wind collision in a
massive binary system. The observed changes can be explained by the
variations in the emission measure of the hot plasma, and by the
different absorption column along the binary orbit. The X-ray spectra
of \wr\ display prominent emission features at wavelengths
corresponding to the lines of strongly ionized Fe, Ca, Ar, S, Si, and
Mg.  WR\,65 is a carbon rich WC9d star that is a persistent dust
maker. This is the first investigation of any X-ray spectrum for a
star of this spectral type.  There are indications that the dust and
the complex geometry of the colliding wind region are pivotal in
explaining the X-ray properties of \wr.

\end{abstract}



\section{Introduction}

{\changed More than 54\% } of all Wolf-Rayet (WR) stars are in binary
systems that consist of a WR and an OB-type star {\changed
\citep{wal07}}. The collision between two stellar winds in these
systems produces characteristic signatures in different wavelength
bands that {\changed can} include non-thermal radio emission
\citep{ei93}, copious and variable X-rays \citep{stev92}, and  
dust emission in the infrared (IR) \citep{wil87}. 

Subject of this {\it Letter} is the WC9d star WR\,65. \citet{wil87}
reported the presence of warm dust in this object. They discussed that
in early-type WC stars an increase in the wind density provided by
{\changed e.g.}\ shock compression in colliding winds in binary systems may be
sufficient to result in grain formation. \citet{zub98} found no need
to invoke binarity to explain the presence of dust in late WC
stars. He fitted the observed IR spectrum of \wr\ assuming a single
star with stellar mass-loss in {\changed the} form of dust with
$\dot{M_{\rm d}}=5.4\times 10^{-10}\,M_\odot\,{\rm yr}^{-1}$ and
surrounded by a dust shell with an inner radius of
520\,$R_\ast$. \citet{lei97} detected radio emission from \wr, but
could not constrain its spectral index.

In a previous paper \citep{osk03} we demonstrated that single WC type
stars are X-ray quiet and proposed that all X-ray active WC stars must
be in binary systems. X-ray emission from \wr\ was detected and
tentatively explained as a result of wind-wind collision. In this
paper we report {\changed on the X-ray light curve of \wr\ } and changes
in its X-ray spectra that unambiguously confirm its status as a
colliding-wind binary (CWB).

\begin{table}
\caption{Coordinates of \wr\ from different sources}
\label{tab:coor}
\centering
\begin{tabular}{ccccc}
\hline
\hline
  Ref           & {\changed Band}   & RA J2000    & DEC J2000 \\  \hline
Simbad          &  opt.    & 15$^{\rm h}$\,13$^{\rm m}$\,41\fs 68 & 
                   $-59\degr\,11\arcmin\,43\farcs 3$ \\ 
\citet{chap99}  & radio & 15$^{\rm h}$\,13$^{\rm m}$\,41\fs 49 & 
                   $-59\degr\,11\arcmin\,43\farcs 8$ \\
\ch\ HRC-I      & X-ray & 15$^{\rm h}$\,13$^{\rm m}$\,41\fs 76 & 
                   $-59\degr\,11\arcmin\,44\farcs 1$ \\            
Catalog USNO-B1.0 & opt.& 15$^{\rm h}$\,13$^{\rm m}$\,41\fs 73 & 
                   $-59\degr\,11\arcmin\,43\farcs 4$ \\ 
\hline
\end{tabular}
\end{table}


\section{WR\,65 and its X-ray light-curve }

The coordinates of \wr\ from different sources are compiled in
Table\,\ref{tab:coor}.  The coordinates from Simbad and the USNO-B1.0
catalog agree well with the source location in the \ch\ HRI-I
image. Except {\changed for} \wr\ there are no other known objects at this
position. WR\,65 is a probable member of the cluster Pismis\,20, which
contains one more WR star, WR\,67, and has a distance of $d=3272\pm
303$\,pc \citep{tur96}.

%
\begin{figure}
\centering
\includegraphics[width=0.7\columnwidth]{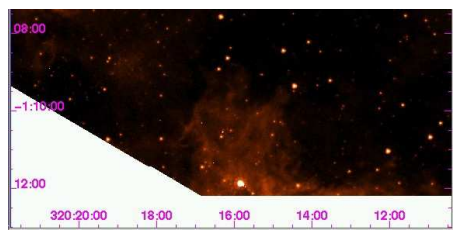}
\caption{Archival {\em Spitzer} image (8\,$\mu$m).  \wr\ is the bright 
object at the bottom, close to the {\changed detector} edge. The 
coordinates are galactic.}
\label{fig:irac}
\end{figure}

We fit the optical spectrum of \wr\ and its available photometry using
the Potsdam Wolf-Rayet (PoWR) stellar atmosphere models
\citep{ham03}. The deduced stellar parameters are listed in
Table\,\ref{tab:par}.

A wind-blown bubble with a radius of $7'$ is known around \wr\ 
\citep{mar97}. \citet{gd04} discovered a H\,{\sc i} shell  with a 
radius of 22\,pc surrounding \wr\ and WR\,67.  The infrared {\em
Spitzer} IRAC image of \wr\ at $8\,\mu$m is shown in
Fig.\,\ref{fig:irac}. Despite its unfavorable location close to the
edge of the {\changed detector}, a rather compact nebula around \wr\
is visible. The radius of the nebula is $\approx 1.5'$ or
1.4\,pc. Given the strong mass-loss rate of \wr\ and its strong UV
field, it is plausible that the nebula is physically associated with
this star.

\begin{table}
\caption{Stellar parameters of \wr\ {\changed (from Hamann et al.,\ in prep.)}}
\label{tab:par}
\centering
\begin{tabular}{cccc}
\hline
\hline
$E_{\rm B-V}$ & $\log{L_{\rm bol}/L_\odot}^{\rm a}$ & $\varv_\infty$ 
[km\,s$^{-1}$] & $\dot{M}\,[M_\odot\,{\rm yr}^{-1}]^{\rm b}$ \\
\hline
2.4   & 5.8\,...\,6.4 & $\approx 2000$ & $10^{-4.6}\,...\,10^{-4.1}$\\
\hline 
\multicolumn{4}{l}{(a)~depending on the adopted reddening law}\\
\multicolumn{4}{l}{(b)~assuming a clumping contrast $D=10$}\\
\end{tabular}
\end{table}

\begin{figure}
\centering
\includegraphics[width=0.7\columnwidth]{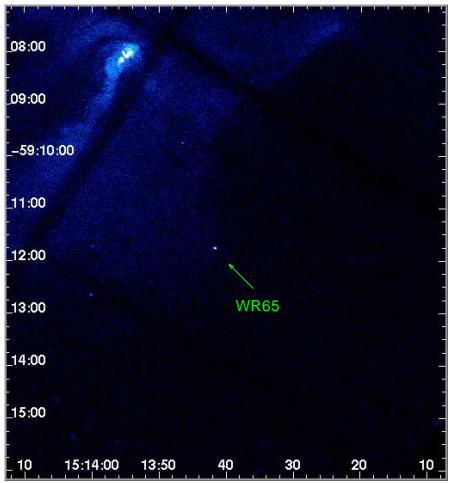}
\caption{Archival \ch\ ACIS-I image (0.2-12.0\,keV) of \wr. The 
observation is from 18\,Oct.\,2005.  The bright object in the 
upper-left corner of the image is the pulsar PSR B1509-58.
\wr\ is projected onto the outskirts of the pulsar wind emission. 
{\changed The coordinates are equatorial (J2000)}. North up, 
east left.}
\label{fig:acis}
\end{figure}

In the X-ray sky, \wr\ is located in a very interesting neighbourhood
(see Fig\,\ref{fig:acis}). The star is $7\farcm 4$ away from the
supernova remnant (SNR) G320.4-1.2 and $4'$ away from the X-ray bright
``Cir Pulsar'' and its associated pulsar wind nebula (PWN). The SNR
and the Cir pulsar are at the distance $d=5.2\pm 1.4$\,kpc
\citep{gen02}.  This region of the sky is often observed, and
serendipitous X-ray observations of \wr\ are in archives.

%

The most homogeneous data set found in archives is from {\em Chandra}
ACIS-I observations conducted from year 2000 to 2005. {\changed In
2005} \wr\ was also observed on one occasion by {\em Chandra} HRC-I
and twice by {\em XMM-Newton}, but was not detected during {\em
XMM-Newton} observations affected by high radiative background.

The local X-ray background is high because \wr\ is projected on the
peripheral region of the PWN mentioned above. Therefore we paid
special attention to the background subtraction.{\changed Conveniently, 
the X-ray emission from the PWN is well studied \citep{gen02,y05}. 
\citet{del06} find that it is constant over at least twelve years. 
We defined background regions as annuli around a stellar point source.}

\begin{figure}
\centering
\includegraphics[width=0.85\columnwidth]{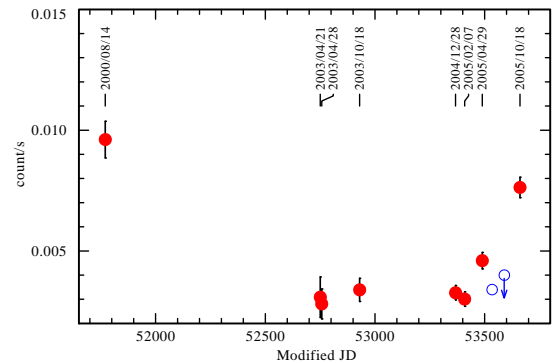}
\caption{X-ray (0.4-10.0\,keV) light-curve of WR\,65. The red dots show 
\ch\ ACIS-I count-rates. The blue dots show count-rates converted to 
ACIS-I rates from  {\em Chandra} HRC-I and {\em XMM-Newton} MOS 
(upper limit).}
\label{fig:lc}
\end{figure}

Figure\,\ref{fig:lc} shows the background-subtracted X-ray light-curve
of \wr\ that displays a surprisingly high level of variability. Such
strong X-ray variability in massive stars is observed only in binary
systems {\changed \citep{cor05,osk01}}. The \wr\ X-ray light curve
gives strong evidence that the star is a colliding wind binary (CWB).
Unfortunately, the light-curve is too sparse to search for the period.
Given the time interval between the two recorded maxima, one may
conclude that the period is not longer than about five years.

\section{WR\,65: X-ray spectral variability}

\begin{figure}
\centering
\includegraphics[width=0.95\columnwidth]{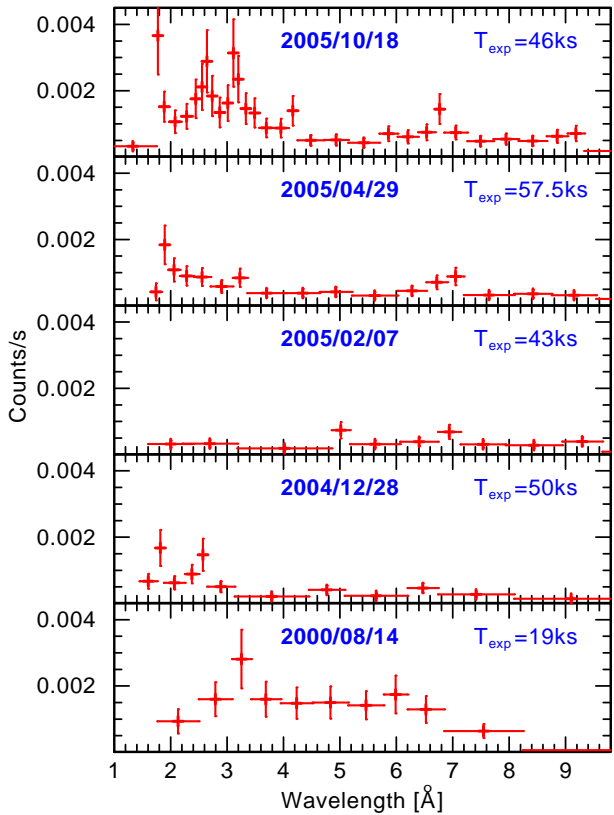}
\caption{{\em Chandra} ACIS-I background subtracted spectra of WR\,65 
at different epochs. The date of the observation and the exposure 
time are given in each panel.}
\label{fig:time}
\end{figure}

\begin{figure}
\centering
\includegraphics[width=0.8\columnwidth]{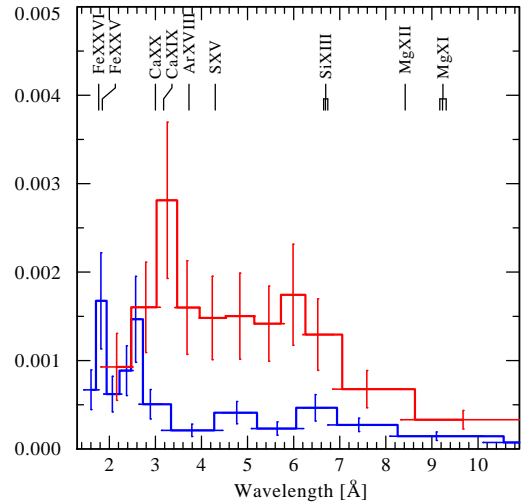}
\caption{{\em Chandra} ACIS-I  spectra of \wr\ in high (red) and 
low (blue) states (obtained on 2000/08/14/ and 2004/12/28
respectively).  Wavelengths of strong spectral lines are indicated.}
\label{fig:2sp}
\end{figure}

Five of the \ch\ observations of \wr\ yielded enough counts for a
crude X-ray spectral analysis. The background-subtracted spectra
obtained at different epochs are displayed in Fig.\,\ref{fig:time}.
The spectral energy distribution is harder than for single WR stars
\citep{ri}. The unresolved emission features coincide with the location 
of lines of Fe, Ca, Ar, S, Si, and Mg (cf.\ Figs.\ref{fig:time},
\ref{fig:2sp}) in collisionally ionized X-ray spectra. Interestingly, 
the iron lines at $\approx 6.7$\,keV {\changed (1.8\,\AA)} are weaker 
at some epochs compared to others (e.g.\ they seem to disappear 
completely on 2005 February 07). {\changed If real, this may 
reflect the change of the temperature in the hot plasma}.

The X-ray spectrum of \wr\ in high states displays strong
emission features in the 2\,-\,4\,\AA\ range (see top panel in
Fig.\,\ref{fig:time}).  Weak lines of Ca\,{\sc xix}\,($\lambda\,3.2$\AA), 
Ca\,{\sc xx}\,($\lambda\,3.0$\AA), and Ar\,{\sc xviii}\,($\lambda\,3.7$\AA) 
are sometimes observed in the X-ray spectra of CWBs, e.g.\ the WC8 binary 
$\gamma^2$\,Vel \citep{sch04}, the WN6
binary WR\,25 \citep{wr25}, and the LBV binary $\eta$\,Car
\citep{hama07}. It seems plausible to assume that the unresolved Ca and
Ar lines contribute to the emission ``bump'' at 2\,-\,4\,\AA\ in the
\wr\ spectrum. However, the strength of this emission complex in \wr\
is much higher than observed in any other CWB. {\changed We rule out 
its origin from the background PWN emission, because there are 
no emission features at 2\,-\,4\,\AA\ in the spectra extracted
from an annulus region around \wr.}

%

In colliding wind binaries consisting of an OB star and a WC star, the
latter has a significantly denser, C, O-enriched wind that is
basically opaque for X-rays \citep{osk03}. When in the course of the
orbital motion the wind collision region (WCR) {\changed is at
superior conjunction relative to the WC star (with respect to the
observer)}, the X-rays generated in the WCR suffer strong absorption.
Despite many still unclear details, this general picture is
observationally confirmed by the X-ray spectral analysis of the WC+O
type binaries WR\,140 and $\gamma^2$\,Vel. In these systems the
strongest photo-attenuation is seen in low X-ray states, while the
lowest photo-attenuation is found when the X-ray flux is high
\citep{gr00,sch04,pol05}.

To investigate how X-ray spectral variations are coupled with the
variation of the absorption column, we fit a model to the observed
spectra and check how the model parameters vary between the different
epochs.

\begin{figure}
\centering
\includegraphics[width=0.55\columnwidth, angle=-90]{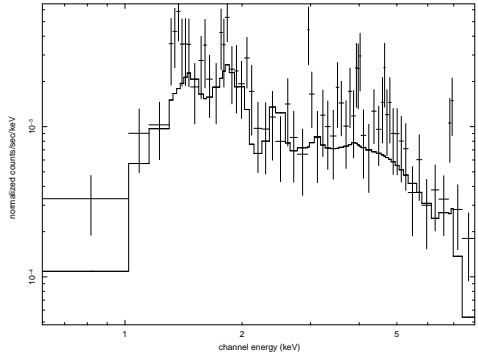}
\caption{ACIS-I spectrum of \wr\ obtained on 2005/10/18 and the 
best fit spectral model -- {\it tbabs}$(N_{\rm H,1})*${\it
apec}$(kT_1)+${\it tbabs}$(N_{\rm H,2})*$ {\it apec}$(kT_2)$. The {\sc
xspec} spectral fitting package was used.  For the model parameters
see text.}
\label{fig:fit}
\end{figure}

\begin{figure}
\centering
\includegraphics[width=0.55\columnwidth, angle=-90]{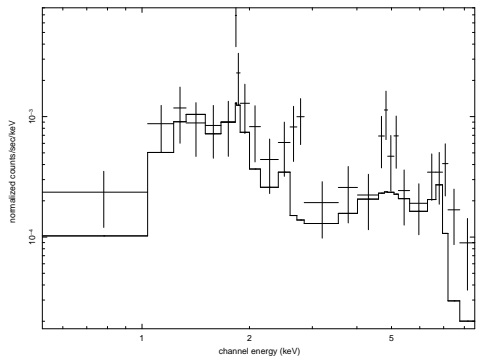}
\caption{Same as in Fig.\,\ref{fig:fit} for the spectrum obtained 
on 2004/12/28.}
\label{fig:fit1}
\end{figure}

Following a method used in \citet{pol05} to successfully analize X-ray
spectra of WR\,140, we tried to fit the observed \wr\ spectra
alternatively with non-equilibrium ionization, {\sc pshock}, and
power-law models using the {\sc xspec} software {\changed
\citep{xspec}}. However, no meaningful constraints on model parameters
could be obtained because of the poor quality of the \wr\ spectra.
\citet{zh07} found that only models with non-equilibrium ionization
can reproduce the {\em detailed} X-ray spectra of CWBs. However, he
noticed that these spectra can be crudely described by thermal plasma
emission with some ``average'' temperature.  Finally, we select a
two-temperature thermal plasma model \citep[{\sc apec}]{apec} 
where each component is allowed to suffer different degrees of attenuation 
-- the ``2T2$N_{\rm H}$'' model. This model provides a first insight into 
the properties of the system by constraining the model parameters.  
We refer to the plasma with lower temperature $kT_1$ as {\changed the} 
"soft" model component, and to the plasma with higher temperature $kT_2$ 
as {\changed the} "hard" model component.  As demonstrated in
Figs.\,\ref{fig:fit},\,\ref{fig:fit1} the 2T2$N_{\rm H}$-model
sufficiently well reproduces the shape of the \wr\ spectra observed at
different epochs.

%

We assume solar composition.  Selecting WC-wind abundances results in
scaling down of the parameter that describes normalization of the
emission measure (EM), but the effect of abundances on {\em spectral
shape} cannot be discriminated given the poor quality of available
data.

The spectrum of \wr\ with highest signal-to-noise and its best-fit
model are shown in Fig.\,\ref{fig:fit}.  The temperatures inferred
from spectral fitting are $kT_1= 1.0\pm 0.2$\,keV and $kT_2= 13.4\pm
7.3$\,keV.  Such temperatures can be expected from the collision of
metal enriched winds in wide binaries. {\changed The  temperature is 
highest at the point of head-on collision, and is decreasing along the 
shock cone} \citep{stev92}.  

The interstellar absorption column density towards the Cir pulsar, 
$N_{\rm H,ism}=1 \times 10^{22}$\,cm$^{-2}$, is well constrained and 
is consistent with the color excess inferred from our fitting of the
\wr\ spectral energy distribution.  The neutral hydrogen
column densities are $N_{H,1}=(2.3\pm 0.3)\times 10^{22}$\,cm$^{-2}$
for the soft component and $N_{H,2}=(12.2\pm 4.1)\times
10^{22}$\,cm$^{-2}$ for the hard component. The column density to the
soft component, $N_{\rm H,1}$, exceeds the interstellar column by a
factor of more than two, but it is lower than the column density to
the hard component, $N_{\rm H,2}$.  This indicates that the hotter
plasma is more deeply embedded.  

From the spectrum obtained at 2005/10/18 the model flux is $F_{\rm
X}(0.6-9.0\,{\rm keV})=1.2\times 10^{-13}$\,erg\,s$^{-1}$\,cm$^{-2}$,
which corresponds to $L_{\rm X}=1.1\times 10^{32}$\,erg\,s$^{-1}$, or
in terms of stellar bolometric luminosity to $\log{L_{\rm X}/L_{\rm
bol}}=-7.2\,...\,-7.8$.  The flux of the soft spectral component,
$F_{\rm X}(0.6-2.0\,{\rm keV})=8.3\times
10^{-15}$\,erg\,s$^{-1}$\,cm$^{-2}$, corresponds to $L_{\rm
X}(0.6-2.0\,{\rm keV})= 1.0 \times 10^{31}$\,erg\,s$^{-1}$ which is
similar to the X-ray luminosity of a single O dwarf \citep{osk05}.

%

%

\begin{figure}
\centering
\includegraphics[width=0.90\columnwidth]{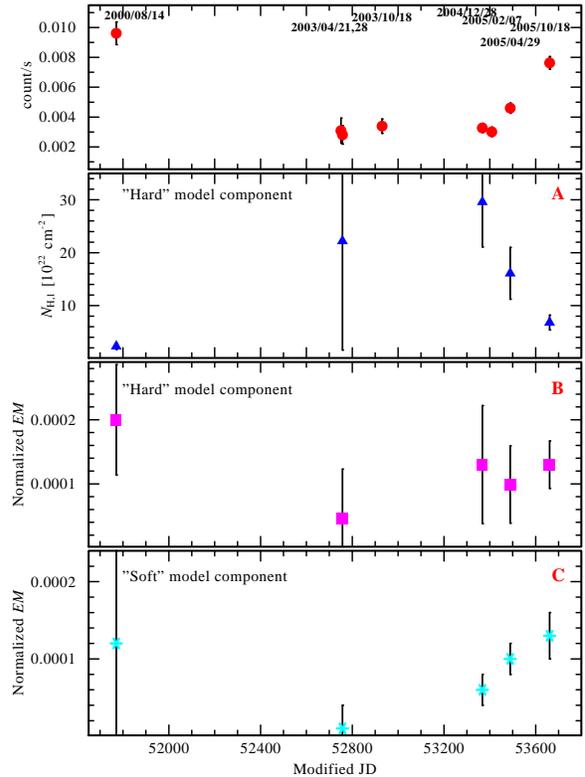}
\caption{{\it Upper panel} same as in Fig.\,\ref{fig:lc}. 
{\it Panel A:} variations of absorption column density of the hard
component  obtained from
spectral fits.  {\it Panel B:} normalization factor that
characterizes the emission measure of the hard component.
 {\it Panel C:} normalization factor of the soft component. }
\label{fig:nh}
\end{figure}

Spectra observed at other epochs were also fitted with the 2T2$N_{\rm
H}$-model.  For illustration, the spectrum obtained in {\changed the}
low state and its corresponding best-fit model are shown in
Fig.\,\ref{fig:fit1}.  Figure\,\ref{fig:nh} shows the model parameters
depending on the time of observation. Within the error bars, the
temperatures $kT_1$ and $kT_2$, and the absorption column to the
source for the soft component, $N_{\rm H,1}$ are similar at all
observed epochs, and therefore are not plotted in
Fig.\,\ref{fig:nh}. In contrast, the column density to the hard
component, $N_{\rm H,2}$, changes at different epochs (panel A of
Fig.\,\ref{fig:nh}).  Variations in the EM of the hard and soft
components are shown in panels B and C of Fig.\,\ref{fig:nh}. The EM
of the soft component changes strongly, reflecting the X-ray
light-curve variations, while the changes in the EM of the hard
component are less pronounced.

%


\section{Discussion \label{seq:diss}}

The main findings from the analysis of \wr\ X-ray spectra obtained at
different epochs are: (i) the spectra display emission features at the
wavelengths of the lines of strongly ionized Fe, Ca, Ar, S, Si, and
Mg; (ii) the star shows strong spectral X-ray variability; (iii) the
spectra can be fitted with a two-temperature model, consisting of a
soft component with $0.3 \lsim kT_1 \lsim 1.5 $\,keV and a hard
component with $2\lsim kT_2 \lsim 20 $\,keV; (iv) the absorption for
the hard component is strong and variable, while the absorption for
the soft component is less severe and nearly constant; (v) the changes
in the EM are stronger for the soft component.

Can these observations be explained in the framework of a colliding
wind model? \citet{cor05} compare the 2.0-10.0\,keV light curves of
WR\,140 and $\eta$ Car (both are systems with high eccentricity and
high inclination) and note that they are very similar. 
\citet{pit98} explain the light-curve (2.0\,--\,10.0\,keV) of 
$\eta$\,Car using a hydrodynamic simulation of colliding winds.  They
show that the data are broadly consistent with the X-ray emission
being moderately low throughout most of the orbit. However, a sharp
peak occurs near periastron, when a greater fraction of the stellar
winds collides, and the higher pre-shock densities and the lower
pre-shock velocities result in stronger radiative cooling. The eclipse
occurs immediately after periastron passage when the companion of
$\eta$\,Car swings behind the LBV star with its dense wind.

It is hard to see how a similar model can be used to explain the X-ray
variability of WR\,65.  Strong variations in the absorbing column
indicate that the orbit of \wr\ has high inclination too. Considering 
the light-curve (Fig.\,\ref{fig:lc}) we can rule out that
\wr\ has a period of a few days: the two data points obtained in April
2003 indicate constant count rates over at least a week. Moreover, two
data points in the light-curve separated by two months (12/2004 and
02/2005) are similar, while there is an increase of emission by nearly
70\% in the next two months. Therefore, \wr\ has most likely an
orbital period of a few years.  In this case the long duration of the
minimum is inconsistent with its interpretation as attenuation by the
wind of the WC star immediately after periastron passage. Quite
different from the short-duration rise of column density observed in
WR\,140 and $\eta$ Car, the column density in \wr\ is persistently
very high, $N_{\rm H,2}\msim 10^{23}\,{\rm cm}^{-2}$, sharply
decreasing only at the maximum X-ray light. 

{\changed The temporary variations of column density in \wr\ are
similar to those measured in the  WC8+O binary $\gamma^2$\,Vel
\citep{gr00}.  Its hard X-ray emission increases by a factor of $\sim 4$
when column density decreases by factor of $\sim$few during the
passage of weak-wind inner side of the shock cone across the line of
sight \citep{aw95,gr00,sch04}.}

In \wr\ the soft X-ray emission is present at all observed epochs. Its
variation can be explained by changing emission measure, while the
absorption is nearly constant and comparably low.  A soft spectral
component in the low state is also observed in $\gamma^2$\,Vel,
V444\,Cyg and $\eta$\,Car.  The origin of soft X-rays in these stars
is attributed either to individual stellar winds or to a circumstellar
nebula. Accordingly, in these objects the absorbing column to the soft
component is similar to the interstellar absorbing column. In
contrast, we find for \wr\ that $N_{\rm H, ism}\,<\,N_{\rm
H,1}\,<\,N_{\rm H, 2}$, and the EM of the soft component changes with
the X-ray light-curve.  Thus it is unlikely that the intrinsic wind
emission of a companion is solely responsible for the soft X-rays from
WR\,65.

WR\,65 is a persistent dust maker \citep{vdh}. \citet{mon02} notice
that binary stars that are persistent dust makers have {\changed
nearly} circular orbits. We can assume that the same holds for
\wr. 
There is a growing number of WC9d binary stars that constitute the
``pinwheel'' class of objects \citep{tut08, mar07}. In these binary
systems the dust is formed in the wind collision region and is
distributed along a wide (compared to the orbital separation)
Archimedean spiral. The peak of dust production occurs at some
distance downstream from the colliding wind stagnation point.

So far there is no direct evidence that \wr\ is a pinwheel. However,
the topology and structure of circumstellar matter in a pinwheel could
help to explain X-ray spectral variability in \wr. \citet{wil00}
presented a model for the absorption of X-rays that includes a
treatment of dust grains. {\changed Their work confirmed that
large-grain absorption for soft X-rays ($< 2$\,keV) is} reduced due
to the self-blanketing of grains.  Interestingly, from the point of
view of radiative transfer this is identical to the reduction of
opacity for the X-rays in clumped stellar winds \citep{feld03}.
{\changed Winds of WC stars are extremely rich in C and O and could
have enhanced abundance of Mg and Ne.}  C and O are dominant absorbers
for the softer X-rays ($< 1$\,keV), while {\changed Mg and Ne
strongly} contribute to the absorption of harder X-rays ($> 1$\,keV).
In the {\changed dusty} shell around a WC9d star, C would be depleted
from the gaseous phase and form grains, reducing the opacity for the
softer X-rays.  This may help to explain the large differences between
the column densities that we determine from soft and hard X-rays.

Recently, \citet{lem07} presented 3-D hydrodynamical simulations of
colliding winds. Scaling the results of their simulation to the
parameters of \wr, soft X-ray emission could originate from a large
volume extending above the orbital plane.  This emission would show
little variation with orbital phase and suffer comparably mild
absorption in the outer regions of the WC star wind, similarly to what
{\changed is} observed in WR\,65. On the other hand, a high postshock
temperature of $kT\approx 10$\,keV is predicted for a confined region
near the point where the winds collide head-on. Densities up to
$\rho\msim 10^{-14}$\,g\,cm$^{-3}$ can be expected in the dusty spiral
arms.  Assuming the wind opacity for hard X-rays as $\kappa\approx
5$\,cm$^2$\,g$^{-1}$ \citep{a04}, a spiral arm width of $10^{14}$\,cm
would explain the observed high absorption.  This width is consistent
with the observed width of the dusty spiral arms in WR\,104
\citep{tut08}. The lowest absorbing column can be expected at the 
phase when the line of sight crosses only the outer low density part
of the spiral arms. It is possible that the \ch\ observations of \wr\
at high state have caught this orbital phase.

To conclude, our first brief study of the variable X-ray emission from
\wr\ indicates that it is a colliding wind system, where the dust and 
complex geometry of its colliding wind region are pivotal in
explaining its X-ray properties. {\changed At present, \wr\ is the only
known WC9d star where X-ray emission is detected.} Future
observations are needed to constrain the orbital parameters of \wr\
and allow for detailed modeling.

\section*{Acknowledgments} The authors are grateful to P. M. Williams for
initiating this study and to N. S. Schulz for useful discussions on
the ACIS-I spectra. {\changed The authors thank the referee 
(A.\,F.\,J. Moffat) for detailed and helpful comments.} Based on
\ch\ data: obsid 754, 6117, 6116, 5535, 5534, 5515, 4384, 3834, 3833.
This research has made use of software provided by the Chandra X-ray
Center and data obtained through the NASA/IPAC Infrared Science
Archive and High Energy Astrophysics Science Archive Research
Center. The SIMBAD database, operated at CDS, Strasbourg, France, and
NASA's Astrophysics Data System were extensively used during this
work.

\label{lastpage}



\begin{thebibliography}{}

\bibitem[\protect\citeauthoryear{Antokhin, Owocki \& Brown}{2004}]{a04}
Antokhin I. I., Owocki S. P., \& Brown J. C., 2004, ApJ, 611, 434 

\bibitem[\protect\citeauthoryear{Arnaud}{1996}]{xspec}
Arnaud K.A., 1996, ASPC, 101, 17

\bibitem[\protect\citeauthoryear{Chapman et al.}{1999}]{chap99}
Chapman J.M., Leitherer C., Koribalski, B., \etal, 1999, ApJ, 518, 890

\bibitem[\protect\citeauthoryear{Corcoran et al.}{2005}]{cor05}
Corcoran M. F., Pittard J. M., Stevens I. R., Henley D. B., Pollock A. M. T.,
2005, in:  X-Ray and Radio Connections, eds. L.O. Sjouwerman and K.K Dyer. 
Published electronically by NRAO, http://www.aoc.nrao.edu/events/xraydio 


\bibitem[\protect\citeauthoryear{DeLaney et al.}{2006}]{del06}
DeLaney T., Gaensler B. M., Arons J., Pivovaroff M. J., 2006,ApJ, 640, 929

\bibitem[\protect\citeauthoryear{Ignace, Oskinova \& Brown}{2003}]{ri}	
Ignace R., Oskinova L. M., Brown J. C., 2003, A\&A, 408, 353

\bibitem[\protect\citeauthoryear{Eichler \& Usov}{1993}]{ei93}
Eichler D., Usov V., 1993, ApJ, 402, 271

\bibitem[\protect\citeauthoryear{Feldmeier, Oskinova \& Hamann}{2003}]{feld03}
Feldmeier A., Oskinova L., Hamann W.-R., 2003, A\&A, 403, 217

	
\bibitem[\protect\citeauthoryear{Gaensler et al.}{2002}]{gen02}
Gaensler B. M., Arons J., Kaspi V. M., Pivovaroff M. J., Kawai N., Tamura K.,
2002, ApJ, 569, 878

\bibitem[\protect\citeauthoryear{Giacani \& Dubner}{2004}]{gd04}
Giacani E., Dubner G., 2004, A\&A, 413, 225

\bibitem[\protect\citeauthoryear{Hamaguchi et al.}{2007}]{hama07}
Hamaguchi K. \etal, 2007, ApJ, 663, 522

\bibitem[\protect\citeauthoryear{Hamann \& Gr{\"a}fener}{2003}]{ham03}
Hamann W.-R., Gr{\"a}fener G., 2003, A\&A, 410, 993.

\bibitem[\protect\citeauthoryear{Leitherer, Chapman \& Koribalski}{1997}]
{lei97} Leitherer C., Chapman J.M., Koribalski B., 1997, ApJ, 481, 898

\bibitem[\protect\citeauthoryear{Lemaster, Stone \& Gardiner}{2007}]{lem07}
Lemaster M. N., Stone J. M.,  Gardiner T. A., 2007, ApJ, 662, 582

\bibitem[\protect\citeauthoryear{Maeda et al.}{1999}]{ma99}
Maeda Y., Koyama K., Yokogawa J., Skinner S., 1999, ApJ, 510, 967

\bibitem[\protect\citeauthoryear{Marchenko \& Moffat}{2007}]{mar07}	
Marchenko S. V., Moffat A. F. J., 2007, ASPC, 367, 213

\bibitem[\protect\citeauthoryear{Marston}{1997}]{mar97}
Marston A. P., 1997, ApJ, 475, 188 

\bibitem[\protect\citeauthoryear{Monnier et al.}{2002}]{mon02}
Monnier J.D., Greenhill L.J., Tuthill P.G., Danchi W.C. 2002, ApJ, 566, 
399 

\bibitem[\protect\citeauthoryear{Oskinova, Clarke \& Pollock}{2001}]{osk01}	
Oskinova L. M., Clarke, D., Pollock, A. M. T., 2001, A\&A, 378, L21

\bibitem[\protect\citeauthoryear{Oskinova et al.}{2003}]{osk03}
Oskinova L. M., Ignace R., Hamann W.-R., Pollock A. M. T., Brown, J. C. 
2003, A\&A, 402, 755

\bibitem[\protect\citeauthoryear{Oskinova}{2005}]{osk05}
Oskinova L. M., 2005, MNRAS, 361, 679

\bibitem[\protect\citeauthoryear{Pittard et al.}{1998}]{pit98}
Pittard J. M., Stevens I. R., Corcoran M. F., Ishibashi, K., 1998, MNRAS, 
299, L5 

\bibitem[\protect\citeauthoryear{Pittard}{2007}]{pit07}
Pittard J. M., 2007, ApJ, 660, L141

\bibitem[\protect\citeauthoryear{Pollock et al.}{2005}]{pol05}	
Pollock A. M. T., Corcoran M. F., Stevens I. R., Williams P. M., 2005, ApJ, 
629, 482

\bibitem[\protect\citeauthoryear{Raassen, van der Hucht \& Mewe}{2003}]{wr25}
Raassen A. J. J., van der Hucht K. A., Mewe R., \etal, 2003, A\&A, 402, 653 

\bibitem[\protect\citeauthoryear{Rauw et al.}{2000}]{gr00}
Rauw G., Stevens I. R., Pittard J. M., Corcoran M. F., 2000, MNRAS, 316, 129

\bibitem[\protect\citeauthoryear{Schild et al.}{2004}]{sch04}
Schild, H., G{\"u}del, M., Mewe, R., \etal\ 2004, A\&A, 422, 177

\bibitem[\protect\citeauthoryear{Smith et al.}{2001}]{apec}
Smith R. K., Brickhouse N. S., Liedahl D. A., Raymond J. C., 2001, ApJ, 556, 
L91

\bibitem[\protect\citeauthoryear{Stevens, Blondin \& Pollock}{1992}]{stev92}
Stevens I. R., Blondin J. M., Pollock A. M. T., 1992, ApJ, 386, 265


\bibitem[\protect\citeauthoryear{Turner}{1996}]{tur96}
Turner D. G., 1996, AJ, 111, 828

\bibitem[\protect\citeauthoryear{Tuthill et al.}{2008}]{tut08}
Tuthill P. G., Monnier J. D., Lawrance N., Danchi W. C., Owocki S. P.,
Gayley K. G., 2008, ApJ, 675, 698

\bibitem[\protect\citeauthoryear{Wallace}{2007}]{wal07}
Wallace D. J., 2007, ASPC, 367, 37

\bibitem[\protect\citeauthoryear{van der Hucht}{2001}]{vdh}
van der Hucht  K. A., 2001, New Astron. Rev., 45, 135 


\bibitem[\protect\citeauthoryear{Williams, van der Hucht \& The}{1987}]{wil87}
Williams P. M., van der Hucht, K. A., Th\'e P. S., 1987, A\&A, 182, 91

\bibitem[\protect\citeauthoryear{Williams et al.}{1990}]{wil90}
Williams P. M., van der Hucht K. A., Pollock A. M. T., \etal, 1990, MNRAS, 
243, 662

\bibitem[\protect\citeauthoryear{Willis, Schild \&  Stevens}{1995}]{aw95}
Willis A. J., Schild H., Stevens, I. R., 1995, A\&A, 298, 549

\bibitem[\protect\citeauthoryear{Wilms, Allen \& McCray}{2000}]{wil00}
Wilms J., Allen A.,  McCray R., 2000, ApJ, 542, 924  

\bibitem[\protect\citeauthoryear{Yatsu et al.}{2005}]{y05}
Yatsu Y., Kawai N., Kataoka J., Kotani T., Tamura K., Brinkmann W., 2005, 
ApJ, 631, 312

\bibitem[\protect\citeauthoryear{Zhekov}{2007}]{zh07}
Zhekov S. A., 2007, MNRAS, 382, 886

\bibitem[\protect\citeauthoryear{Zubko}{1998}]{zub98}
Zubko V. G., 1998, MNRAS, 295, 109


\end{thebibliography}
\end{document}